\documentclass[twoside]{ilcws10}
\usepackage[latin1]{inputenc}
\usepackage[dvips]{graphicx,epsfig,color}
\usepackage{wrapfig,rotating}
\usepackage{amssymb,amsmath,array}

\begin{document}
\title{
%%%%   Paper title goes here  %%%%%%%%%%%%%%
Simulation Study of \boldmath{$W$} Boson + Dark Matter Signatures for Identification of New Physics} %% 
%***********************************************************************
% AUTHORS INFORMATION AREA
%***********************************************************************
\author{T.~Suehara$^1$\thanks{e-mail: suehara@icepp.s.u-tokyo.ac.jp},
M.~Asano$^2$, K.~Fujii$^3$, R.~S.~Hundi$^4$, H.~Itoh$^{3,5}$,\\
S.~Matsumoto$^6$, N.~Okada$^7$, T.~Saito$^2$, Y.~Takubo$^8$, and H.~Yamamoto$^2$
% Optional short acknowledgment: remove next line if non-needed
% DO NOT MODIFY THE FOLLOWING '\vspace' ARGUMENT
\vspace{.3cm}\\
1- The University of Tokyo - International Center for Elementary Particle Physics \\
7-3-1 Hongo, Bunkyo-ku, Tokyo, 113-0033 - Japan
\vspace{.1cm}\\
2- Tohoku University - Department of Physics \\
6-3 Aoba, Aramaki, Aoba-ku, Sendai, Miyagi, 980-8578 - Japan
\vspace{.1cm}\\
3- High Energy Accelerator Research Organization - Institute of Particle and Nuclear Studies \\
1-1 Oho, Tsukuba, Ibaraki, 305-0801 - Japan
\vspace{.1cm}\\
4- University of Hawaii - Department of Physics \& Astronomy \\
2505 Correa Rd.~Honolulu, HI, 96822 - United States
\vspace{.1cm}\\
5- The University of Tokyo - Institute of Cosmic Ray Research \\
5-1-5 Kashiwa-no-ha, Kashiwa, Chiba, 277-8582 - Japan
\vspace{.1cm}\\
6- University of Toyama - Department of Physics, Graduate School of Science and Engineering \\
3190 Gofuku, Toyama, Toyama, 930-8555 - Japan
\vspace{.1cm}\\
7- The University of  Alabama - Department of Physics \& Astronomy \\
Tuscaloosa, AL, 35487-0324 - United States
\vspace{.1cm}\\
8- Tohoku University - Center for the Advancement of Higher Education \\
41 Kawauchi, Aoba-ku, Sendai, Miyagi, 980-8576 - Japan
}
%%***********************************************************************
% END OF AUTHORS INFORMATION AREA
%***********************************************************************

\maketitle

\begin{abstract}
Identification of beyond-standard-models including WIMP dark matter is studied in
four particle final state with a $W$ boson pair and a WIMP pair
at the International Linear Collider.
Models with different spin structures give distinguishable production angle distributions.
After the mass determination in each model, the production angle is reconstructed using
the four momentum of $W$ bosons with a back-to-back constraint. Three models of Inert Higgs, Supersymmetry
and Little Higgs are considered. Discrimination power at 200 fb and 40 fb signal cross section
with 500 fb$^{-1}$ luminosity at $\sqrt{s}=500$ GeV is obtained.
\end{abstract}

\section{Introduction}

A weakly interacting massive particle (WIMP) is a leading candidate for dark matter.
Many models including particles of various spin are proposed in order to explain
existence of the WIMP dark matter.
Using angular distribution, those models may be separated in collider experiments.

In this study, we investigate the separation possibility in the process of
e$^+$e$^-$ $\rightarrow$ $\chi^+\chi^-$ $\rightarrow$ $\chi^0\chi^0$W$^+$W$^-$
at the International Linear Collider (ILC) with full Monte Carlo simulation.
$\chi^0$ and $\chi^\pm$ denote the WIMP and the lightest charged particle (LCP) in the model, respectively.
Center-of-mass energy of ILC is set to be 500 GeV.
Particles in the model other than the LCP and the WIMP are assumed to
be inaccessible at this energy.

We consider three WIMP model candidates as shown in Table \ref{tbl:models}.
Since the three models have different spins of their particles (Inert Higgs (IH)\cite{ih}: scalar,
Supersymmetry (SUSY)\cite{susy}: fermion, Little Higgs (LH)\cite{lht}: vector), they are good samples
for the model identification study.
Mass is set to be the same as m($\chi^\pm$) = 231.57 GeV and m($\chi^0$) = 44.03 GeV
for each model by tuning its model parameters.
Cross section is normalized to 40 fb and 200 fb for the analysis.

\begin{table}
\centerline{
	\begin{tabular}{|c|c|c|r|r|r|}
	\hline
	Model         & Particle & Name       & Spin & Mass       & $\sigma$ at 500 GeV \\\hline\hline
	Inert Higgs   & LCP      & $\eta^\pm$ &    0 & 231.57 GeV & 3.51 fb \\
	              & WIMP     & $\eta_I$ &    0 & 44.03 GeV & \\\hline
	Supersymmetry & LCP      & $\chi^\pm$ &  1/2 & 231.57 GeV & 384 fb \\
	              & WIMP     & $\chi^0_1$ &  1/2 & 44.03 GeV & \\\hline
	Little Higgs  & LCP      & $W_\mathrm{H}^\pm$ & 1 & 231.57 GeV & 364 fb \\
	              & WIMP     & $A_\mathrm{H}$ & 1 & 44.03 GeV & \\\hline
	\end{tabular}
}
\caption{Models and cross sections used for this study. In the analysis
Cross sections are normalized to 40/200 fb so the cross section listed in this table is not used.}
\label{tbl:models}
\end{table}

\section{Simulation and Reconstruction}

The event generation is performed using Physsim\cite{physsim} event generator for events of three sample models
and Whizard\cite{whizard} generator for the Standard Model (SM) background sample.
The initial-state radiation and beamstrahlung are included in the event generations.
The RDR\cite{ilcrdr} nominal beam parameters are used while the finite crossing angle is ignored.
We assume no beam polarization.

According to the assumed cross section of 200 fb, 0.1 million signal events are generated for each model.
The SM background sample is prepared for International Large Detector (ILD) Letter of Intent (LoI) \cite{ildloi}.
Its statistics corresponds to 0.1 - 500 fb$^{-1}$ luminosity (depends on processes) with about 12 million events in total.

ILD full simulation code (Mokka\cite{mokka} and MarlinReco\cite{marlinreco}) is used for the full Monte-Carlo (MC) simulation
and event reconstruction. ILD\_00 detector geometry, which is the standard geometry for ILD LoI is used for the simulation.
The geometry includes a time projection chamber with silicon devices for tracking and vertexing, and highly granular
 electromagnetic and hadronic calorimeters for particle flow calorimetry along with 3.5 Tesla magnetic field.
The simulation code includes many realistic features, gaps in the sensitive regions
and estimates of dead material due to cables, mechanical support, cooling and so on.
The central part of the reconstruction is the particle flow algorithm PandoraPFA~\cite{pandora}, which forms charged 
and neutral particle candidates (particle flow objects: PFOs) from 
tracks and calorimeter clusters. The resulting list of PFOs for each event is forced 
into a 4-jet configuration using the Durham algorithm\cite{durham}.
Neural-net based flavor tagging algorithm LCFIVertex\cite{lcfi}
is applied for the jets after the clustering and vertex finding with ZVTOP\cite{zvtop} algorithm.
As a final step of the reconstruction, a constrained kinematic fit~\cite{kinfitnote}, which 
requires the two dijet masses of the event to be equal, is performed on each event. All three possible jet pairings are tested
and the pairing with least $\chi^2$ value for the kinematic fit is adopted.

\section{Signal Selection}

We use signal events with both $W$ decaying into two quarks ($qqqq$ events), which has about 46\% branching ratio,
since $W$ energy must be fully reconstructed for the mass determination and the production angle reconstruction.
The signature of the fully hadronic decay from 
our target processes is `4-quarks $+$ missing'.
The main SM background processes are
$W$- and $Z$-pair production with fully hadronic decay,
top-pair production with one leptonic decay of $W$,
$\gamma\gamma \rightarrow WW$ processes, $WWZ$ process with the $Z$ decays to a $\nu$-pair, etc.
Semi-leptonic decay of the target processes is also background for our study.

The following selection is applied to all samples to reject
the major part of SM and semi-leptonic decay background:
(i) Number of tracks should be larger than $20$ and
each jet has to contain at least two tracks in order to eliminate pure leptonic events,
(ii) The visible energy of the event $E_{\mathrm{vis}}$ should be between $80$ and $400$ GeV
which can remove most of 2-photon and 2/4/6 quark events,
(iii) each jet should have a reconstructed energy of at least 5~GeV and a polar angle $\theta$ 
  fulfilling $|\cos(\theta_{\mathrm{jet}})| < 0.99$ to ensure proper jet reconstruction,
(iv) Requiring the distance parameter
  of the Durham jet algorithm\cite{durham} for which the event changes from 4-jet to 3-jet
  configuration, $y_{34}$ to be larger than 0.001 in order to reject most of 2-jet events,
(v) No lepton candidate with an energy larger than 25~GeV is allowed
  in order to suppress semi-leptonic events,
(vi) $|\cos(\theta)|$ of the missing momentum should be smaller than 0.9
and $|\cos(\theta)|$ summed up for all jets should be smaller than $2.6$
in order to eliminate most of SM events which are concentrated in the forward region,
(vii) B-tag probability from LCFIVertex output summed up for all jets should be smaller than 1
to remove events with b quarks.
(viii) The kinematic fit constraining the two dijet masses to be equal should converge for at 
least one jet pairing to ensure integrity of the fit result.
(ix) The di-jet mass obtained by the kinematic fit should be between $65$ and $95$ GeV
to select two-$W$ events.

\begin{table}
\begin{center}
                \begin{tabular}{|c|l|r|r|r|r|} \hline
           & Processes                                              & No cut   & before mass cut & all cuts \\ \hline\hline
           & Inert Higgs (hadronic decay)                           & 46746    & 33422           & 27943    \\ \cline{2-5}
Signal     & SUSY (hadronic decay)                                  & 45539    & 31942           & 26552    \\ \cline{2-5}
           & Little Higgs (hadronic decay)                          & 46466    & 32681           & 27361    \\ \hline\hline
           & Inert Higgs (other decay)                              & 53252    &  1973           &   185    \\ \cline{2-5}
Model bkg. & SUSY (other decay)                                     & 54462    &  1930           &   139    \\ \cline{2-5}
           & Little Higgs (other decay)                             & 53532    &  1770           &   239    \\ \hline\hline
           & $qqqq$ ($WW$, $ZZ$)                                    & 1.88e+06 &  8360           &  3223    \\ \cline{2-5}
           & $qq\ell\nu$ ($WW$)                                     & 2.35e+06 &  6796           &  1885    \\ \cline{2-5}
           & $qqqq\ell\nu$ ($tt$)                                   & 125204   &  2754           &   626    \\ \cline{2-5}
           & $qq\ell\nu\ell\nu$                                     & 31630    &   103           &    39    \\ \cline{2-5}
SM bkg.    & $\gamma\gamma\rightarrow{}qqqq$                        & 26356    &   666           &   523    \\ \cline{2-5}
           & $qqqq\nu\nu$ ($WWZ$)                                   & 4158     &  1843           &   681    \\ \cline{2-5}
           & $qq\nu\nu$ ($ZZ$)                                      & 117808   &  8547           &   134    \\ \cline{2-5}
           & $qq$                                                   & 6.29e+06 &  1518           &   373    \\ \cline{2-5}
           & $\gamma\gamma\rightarrow{}qq$                          & 7.97e+06 &   914           &    58    \\ \cline{2-5}
           & Other background                                       & 3.42e+09 &   727           &   108    \\ \hline
                \end{tabular}
\caption{Event number before and after selection cuts for 200 fb production cross section, normalized to 500 fb$^{-1}$ and no beam polarization.}
\label{tbl:cutstat500}
\end{center}
\end{table}

\begin{figure}[thb]
% 	\begin{minipage}[t]{.24\textwidth}
% 	\begin{center}
% 		\includegraphics[width=0.95\columnwidth]{ntrack.eps}
% 		(i)\\[1cm]
% 	\end{center}
% 	\end{minipage}
	\begin{minipage}[t]{.32\textwidth}
	\begin{center}
		\includegraphics[width=1\columnwidth]{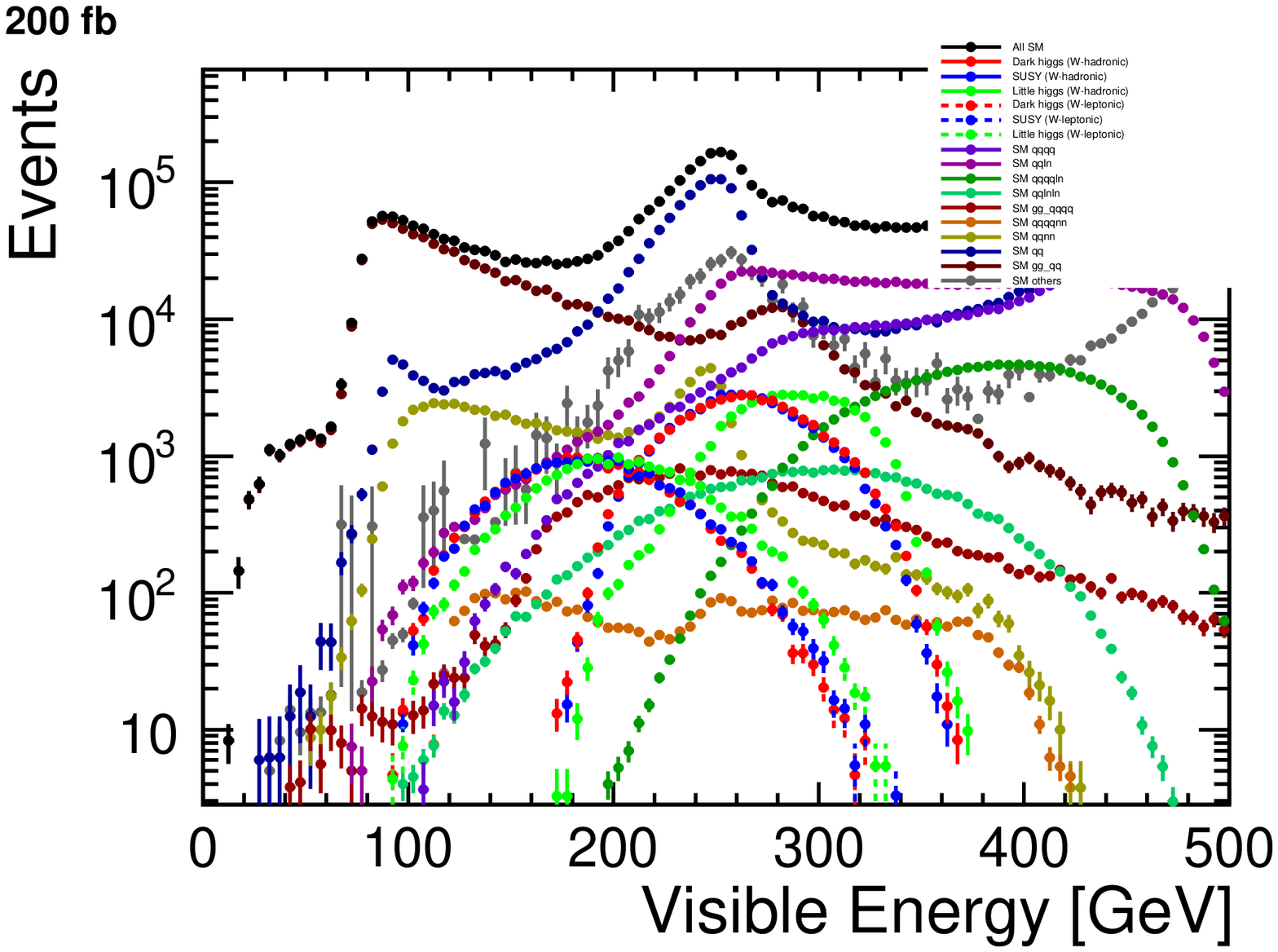}
		(ii)\\[1cm]
	\end{center}
	\end{minipage}
	\begin{minipage}[t]{.32\textwidth}
	\begin{center}
		\includegraphics[width=1\columnwidth]{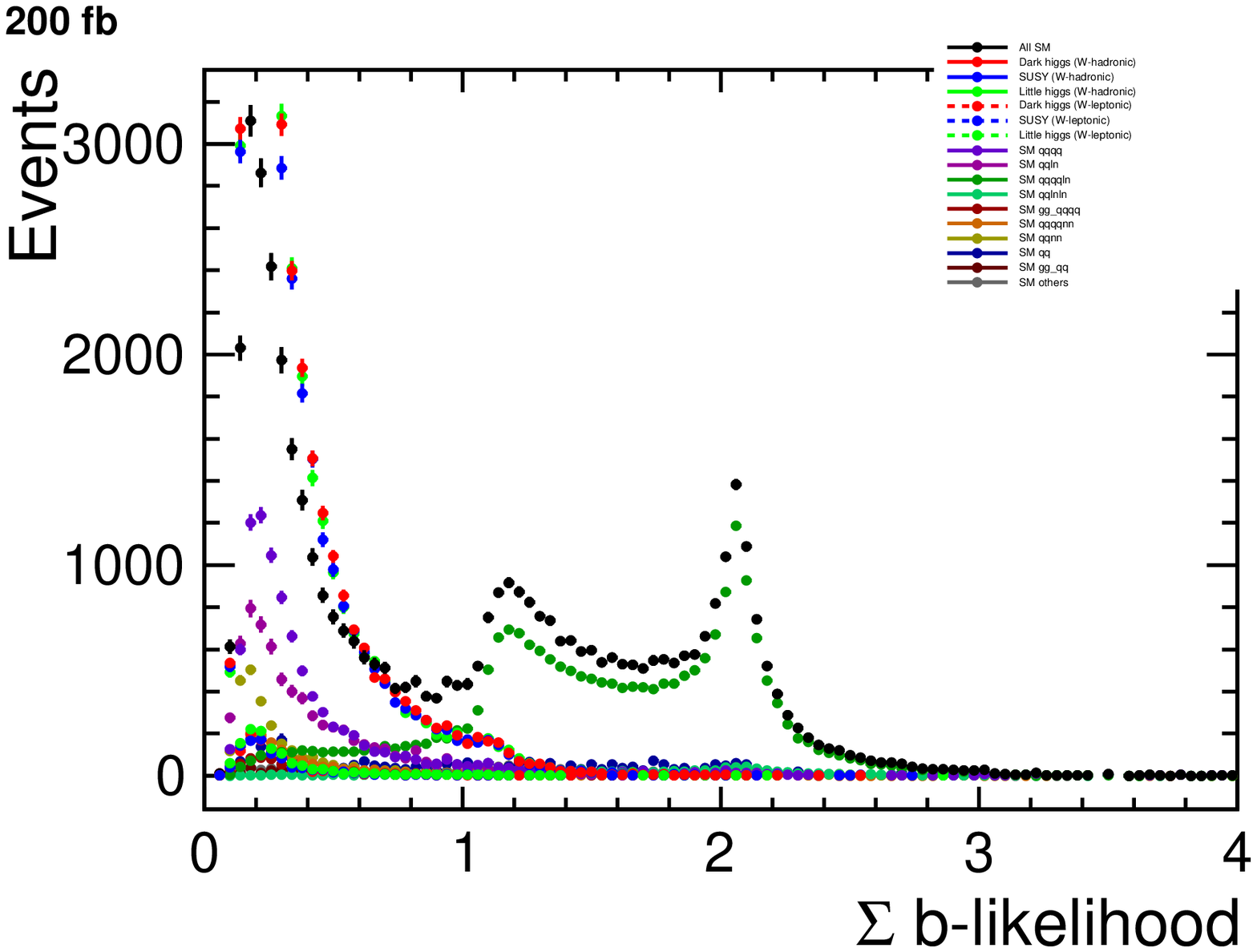}
		(vii)\\[1cm]
	\end{center}
	\end{minipage}
	\begin{minipage}[t]{.32\textwidth}
	\begin{center}
		\includegraphics[width=1\columnwidth]{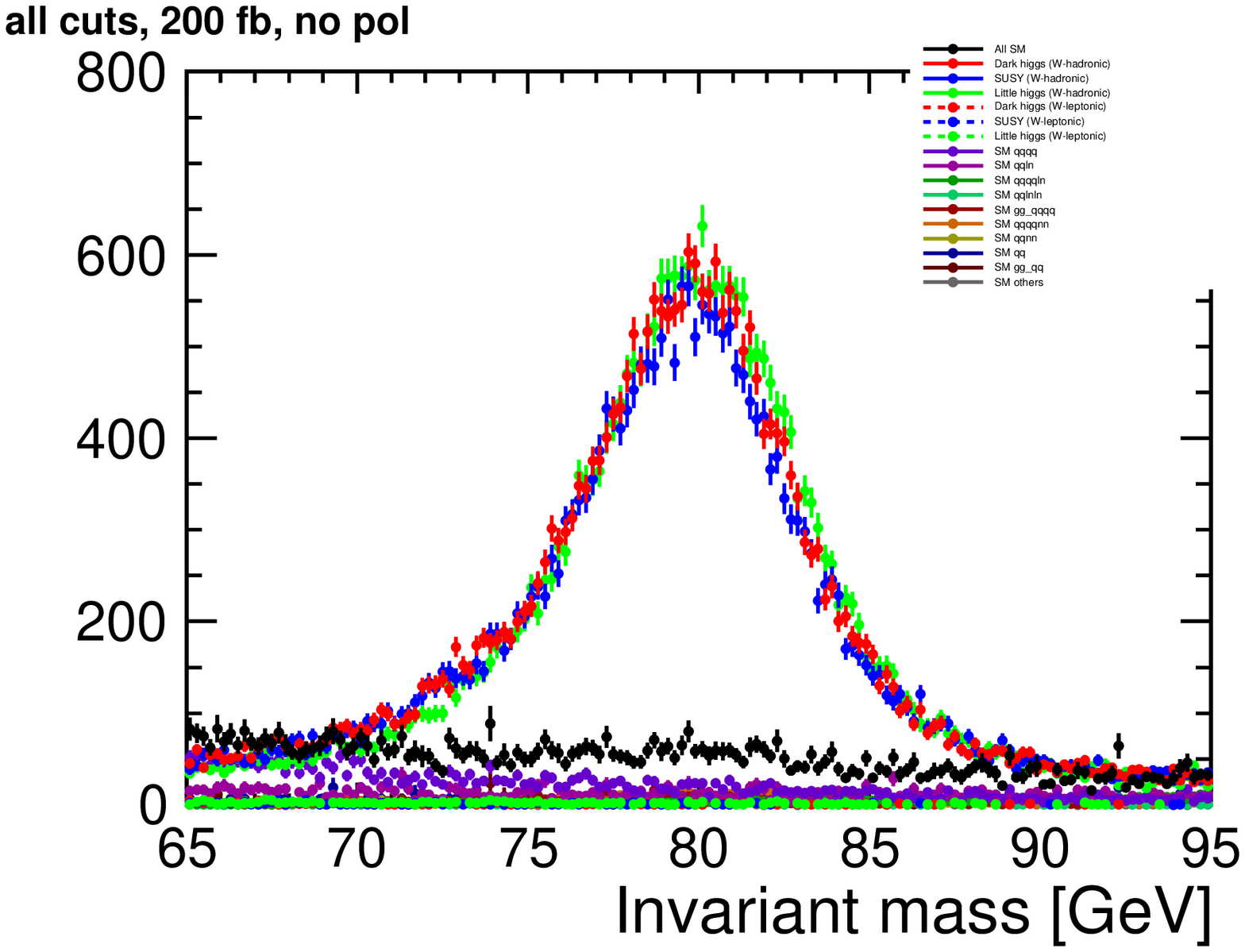}
		(ix)\\[1cm]
	\end{center}
	\end{minipage}
	\caption{Cut plots in 200 fb signal cross section. The labels (ii), (vii), (ix) correspond to
the cuts described in the text with the same labels.}
	\label{fig:cutplot500}
\end{figure}

The obtained reduction due to these cuts is shown in Table~\ref{tbl:cutstat500} and
some of the cut plots are shown in Figure \ref{fig:cutplot500}.
Clear peaks at di-jet masses of 80 GeV can be seen in the signal distributions of 
Figure \ref{fig:cutplot500} (ix), which are from two W bosons.
Acceptances of signal events are 59.8, 58.3 and 58.9 \% for IH, SUSY and LH
models, respectively. Signal purities after the cuts are 78.1, 77.3 and 77.6 \%
in 200 fb cross section and 42.1, 40.9 and 41.6 \% in 40 fb cross section, respectively.

\section{Mass Determination}

\begin{figure}
	\begin{minipage}[t]{.32\textwidth}
	\begin{center}
    \includegraphics[width=1\textwidth]{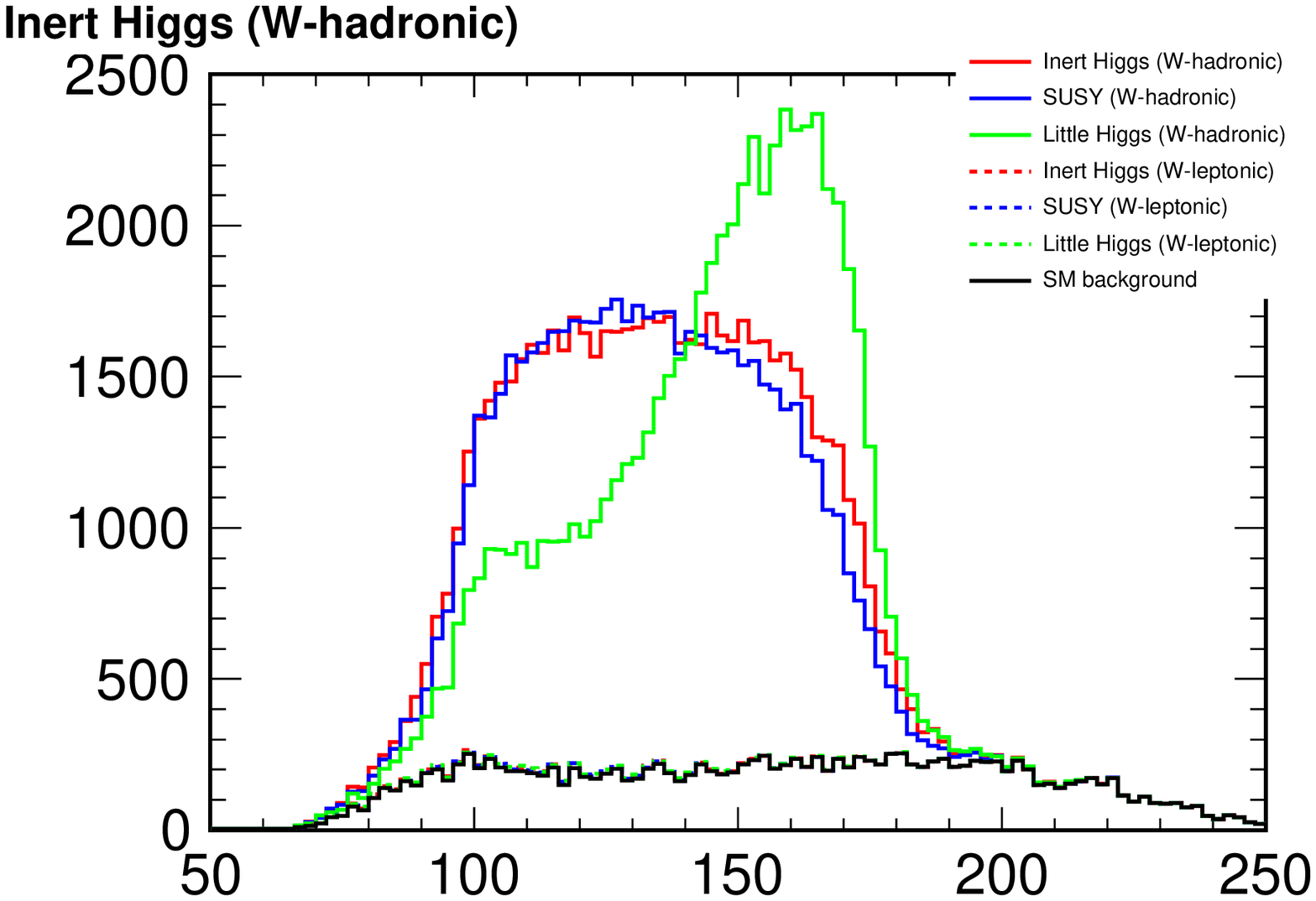}
		(a)
	\end{center}
	\end{minipage}
	\begin{minipage}[t]{.32\textwidth}
	\begin{center}
    \includegraphics[width=1\textwidth]{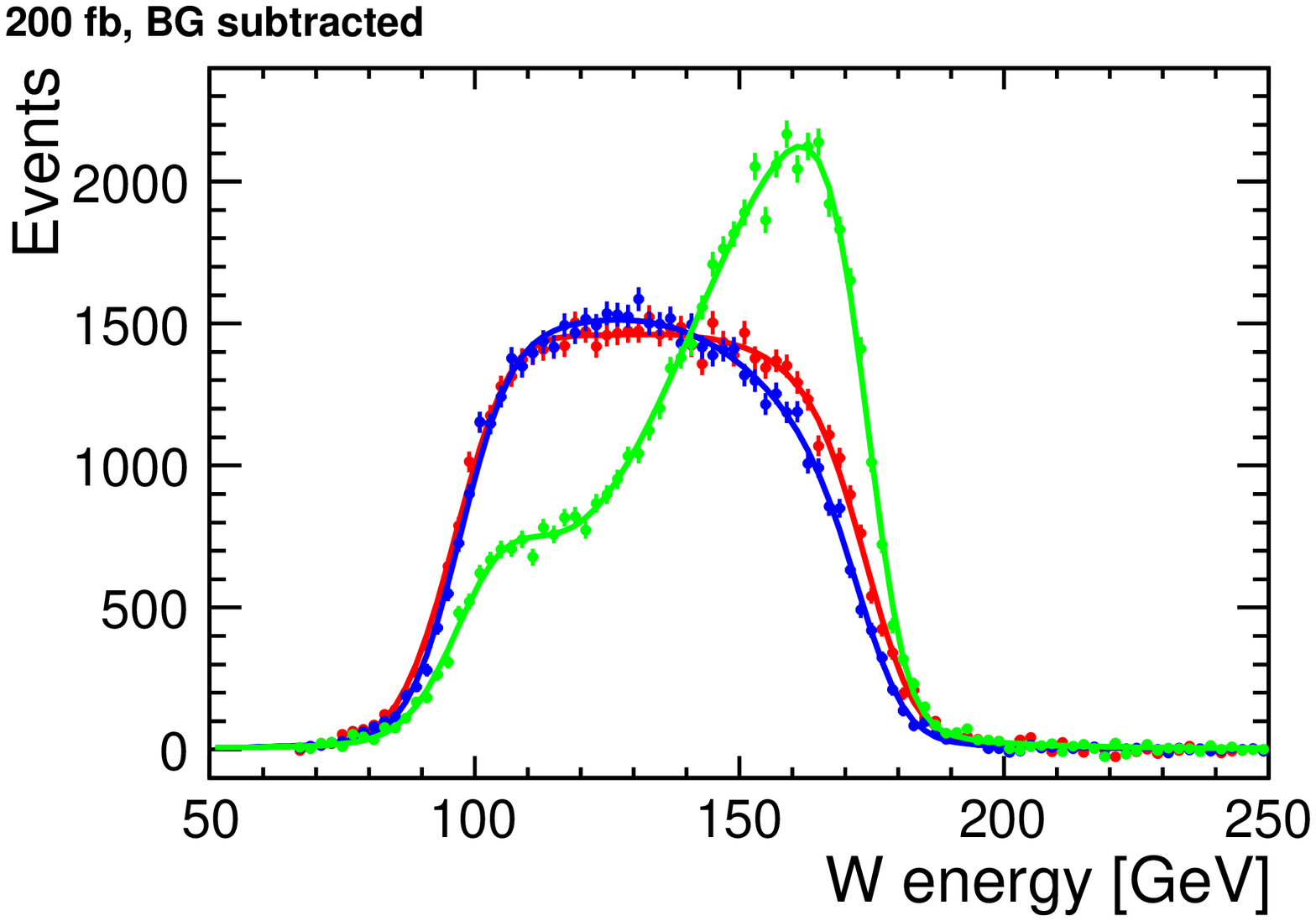}
		(b)
	\end{center}
	\end{minipage}
	\begin{minipage}[t]{.32\textwidth}
	\begin{center}
    \includegraphics[width=1\textwidth]{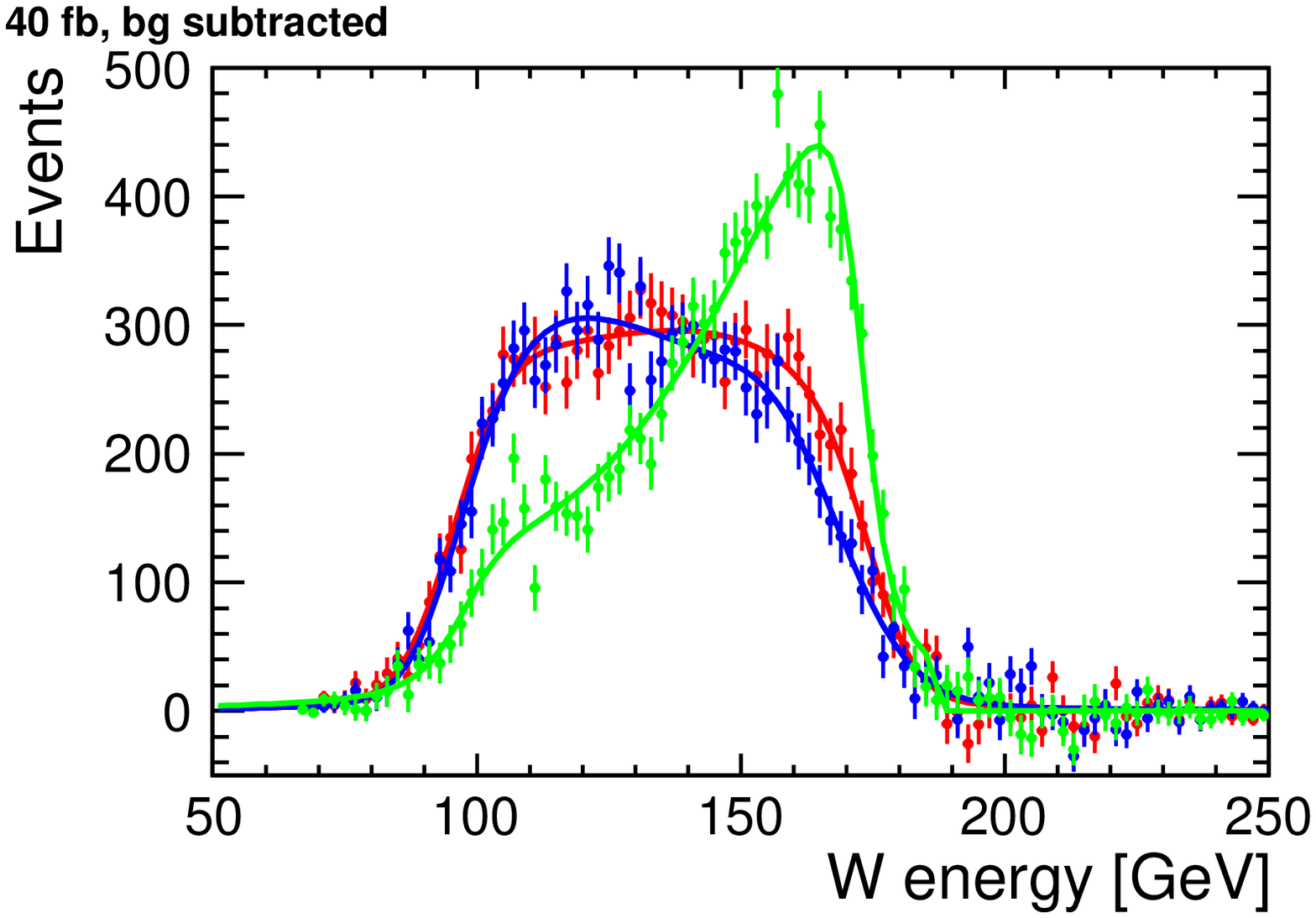}
		(c)
	\end{center}
	\end{minipage}
	\caption{(a)W energy distribution of signal and background.
		(b),(c) Results of the mass fit for 200 fb and 40 fb signal cross section after background subtraction, respectively.}	
	\label{fig:wmass500}
\end{figure}

The masses of new particles can be obtained via the energy spectrum of the $W$ boson candidates.
The energy of $W$ bosons by production of new particles have
upper and lower kinematic limits from which the masses of the new particles can be derived.
Figure \ref{fig:wmass500}(a) shows the $W$ energy spectrum for each model with SM background.
Mass edges can be seen in the distribution of every model.

The edge positions of the energy spectra are obtained by fitting.
The fitting function is:

	\begin{equation}
		f(x;t_{0-1},b_{0-2},\sigma_{0-1},\Gamma) = \int^{t_1}_{t_0}(b_2t^2+b_1t+b_0)V(t-x,\sigma_1{}t + \sigma_0,\Gamma)dt
	\end{equation}

where $V(x, \sigma, \Gamma)$ denotes the Voigt function of Gaussian width of $\sigma$ and Lorenzian width of $\Gamma$.
After the fit with all parameters free (8 free parameters), we fixed all the parameters except $t_0$ and $t_1$ (which gives the edge positions)
 and refit to obtain shape-independent mass resolution.
Figure \ref{fig:wmass500}(b)(c) gives the fitting result with 200 fb and 40 fb signal cross section, respectively.
The fitting result is summarized in Table \ref{tbl:massfit}.
While the center value of the fitting result deviates from the expected true edge position,
it can be corrected using Monte Carlo samples in the real experiment.

\begin{table}
\centerline{
	\begin{tabular}{|c|r|r|r|r|r|}
	\hline
	Model         & $\sigma$ & Fitted left edge     & Fitted right edge     & LCP $\delta{}$m & WIMP $\delta$m \\\hline\hline
	True edges		&					 & 96.32 GeV						& 174.78 GeV						&									&	 \\\hline
	Inert Higgs   &  200 fb  & 97.00 $\pm$ 0.13 GeV & 174.22 $\pm$ 0.19 GeV &				 0.12 GeV & 			0.57 GeV \\
								&   40 fb  & 97.86 $\pm$ 0.38 GeV & 176.22 $\pm$ 0.84 GeV &				 0.39 GeV &				2.23 GeV \\\hline
	Supersymmetry &  200 fb  & 97.12 $\pm$ 0.15 GeV & 173.46 $\pm$ 0.21 GeV & 			 0.14 GeV &				0.64 GeV \\
								&	  40 fb  & 96.10 $\pm$ 0.54 GeV & 176.02 $\pm$ 0.85 GeV & 			 0.50 GeV &				2.50 GeV \\\hline
	Little Higgs  &	 200 fb  & 96.94 $\pm$ 0.16 GeV & 174.39 $\pm$ 0.10 GeV &				 0.13 GeV &				0.50 GeV \\
								&	  40 fb  & 99.56 $\pm$ 0.48 GeV & 174.67 $\pm$ 0.34 GeV &				 0.41 GeV &				1.56 GeV \\\hline
	\end{tabular}
}
\caption{Result of the mass fit. Refer to the text for $\delta{}m$ calculation.}
\label{tbl:massfit}
\end{table}

The masses of LCP and WIMP can be calculated from the edge positions.
Error of the edge positions can be translated to the error of the masses $\delta{}m$.
For the calculation, true edge positions are used instead of the (deviated) center value of the fitting result.
The obtained mass error is also shown in Table \ref{tbl:massfit}.

Since we can see clear difference among three models in the $W$ energy distribution,
the model separation may be possible with the difference.
However, the difference is considered to come from the specific model and not from
the general spin structure, we do not use the difference to identify the model in this study.

\section{Angular Distribution for LCP Pair Production}

\begin{figure}
	\begin{center}
	\begin{minipage}[t]{.35\textwidth}
	\begin{center}
    \includegraphics[width=1\textwidth]{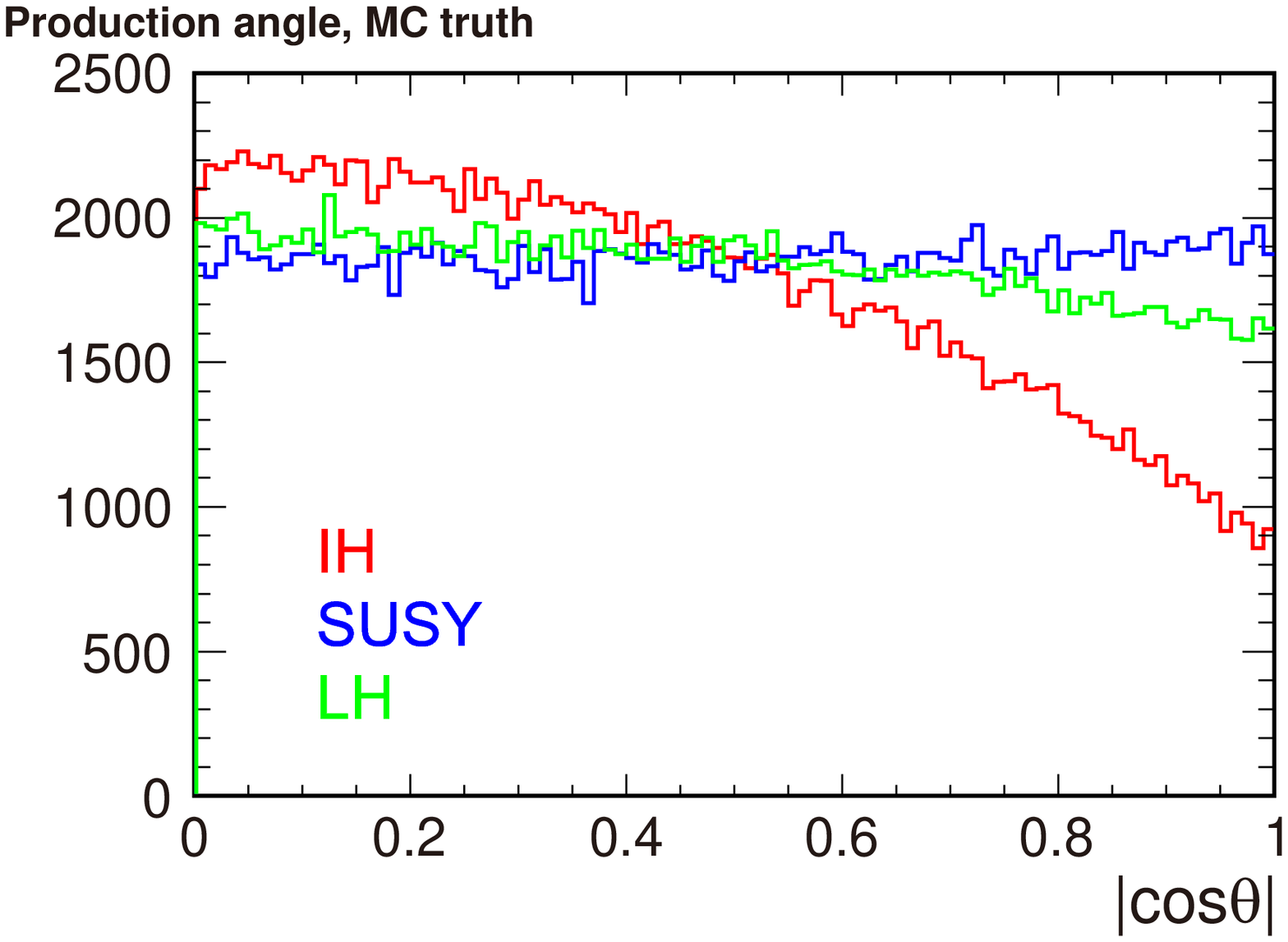}
		(a)
	\end{center}
	\end{minipage}
	\begin{minipage}[t]{.35\textwidth}
	\begin{center}
    \includegraphics[width=1\textwidth]{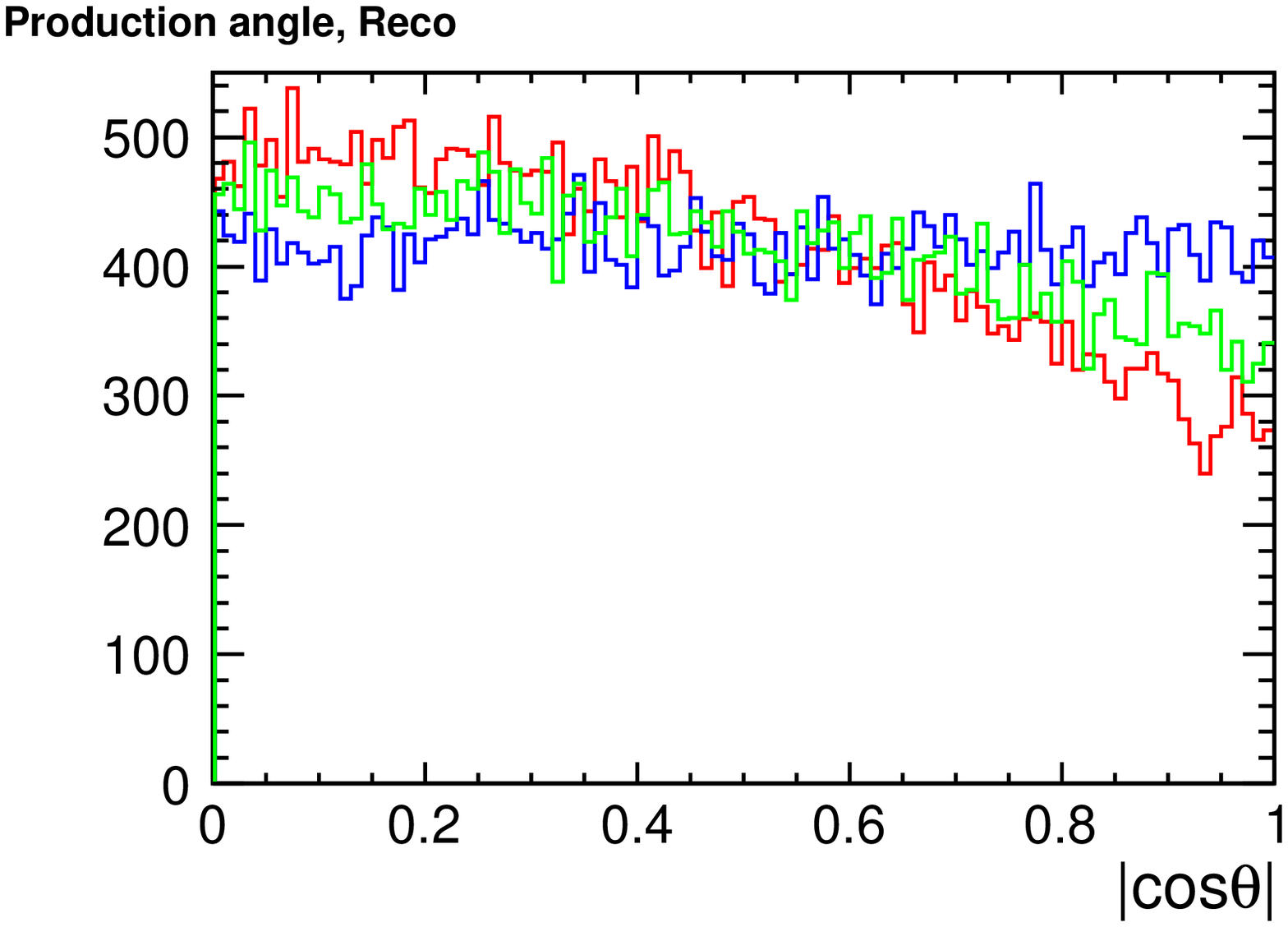}
		(b)
	\end{center}
	\end{minipage}
\\
	\begin{minipage}[t]{.24\textwidth}
	\begin{center}
    \includegraphics[width=1\textwidth]{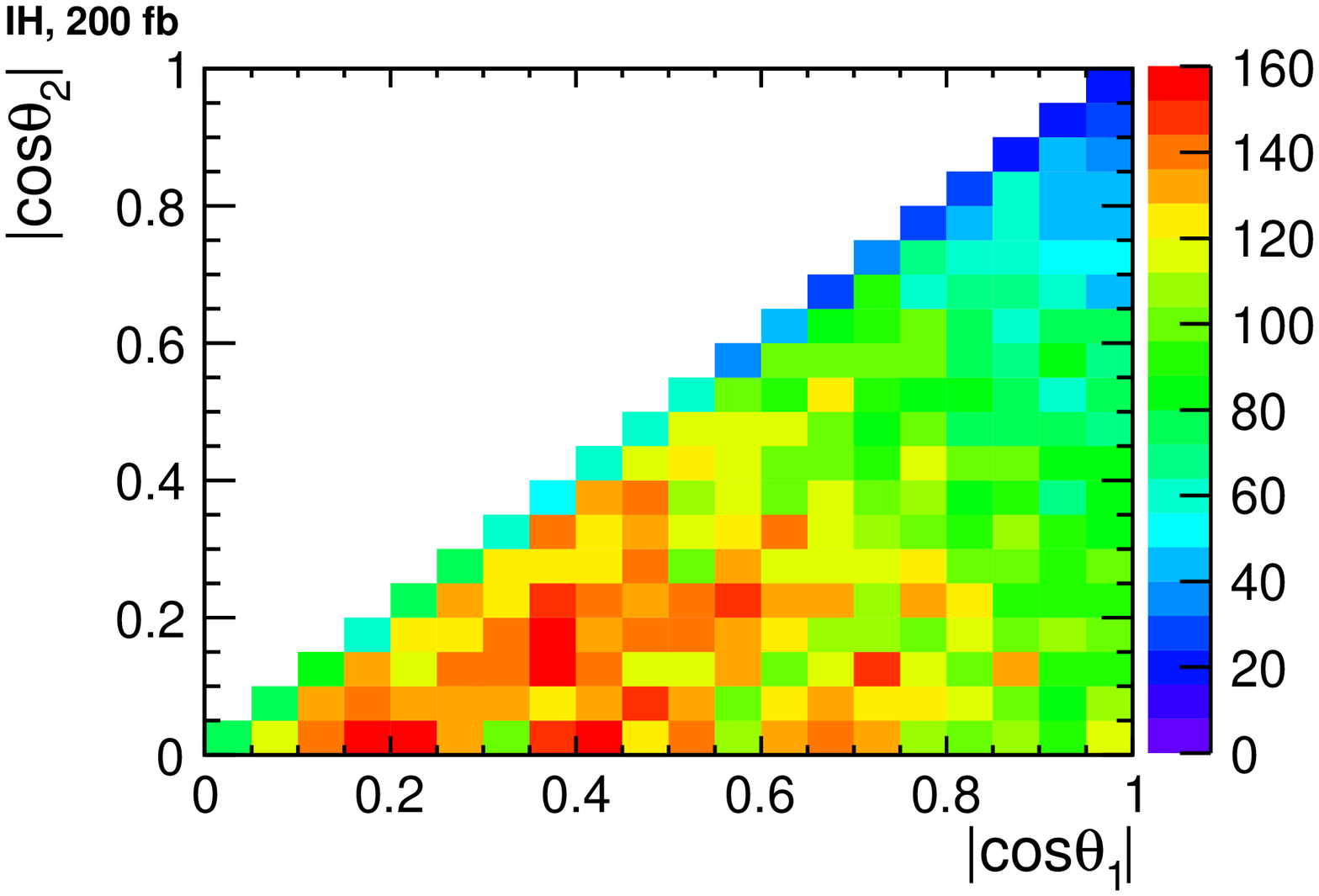}
		(c)
	\end{center}
	\end{minipage}
	\begin{minipage}[t]{.24\textwidth}
	\begin{center}
    \includegraphics[width=1\textwidth]{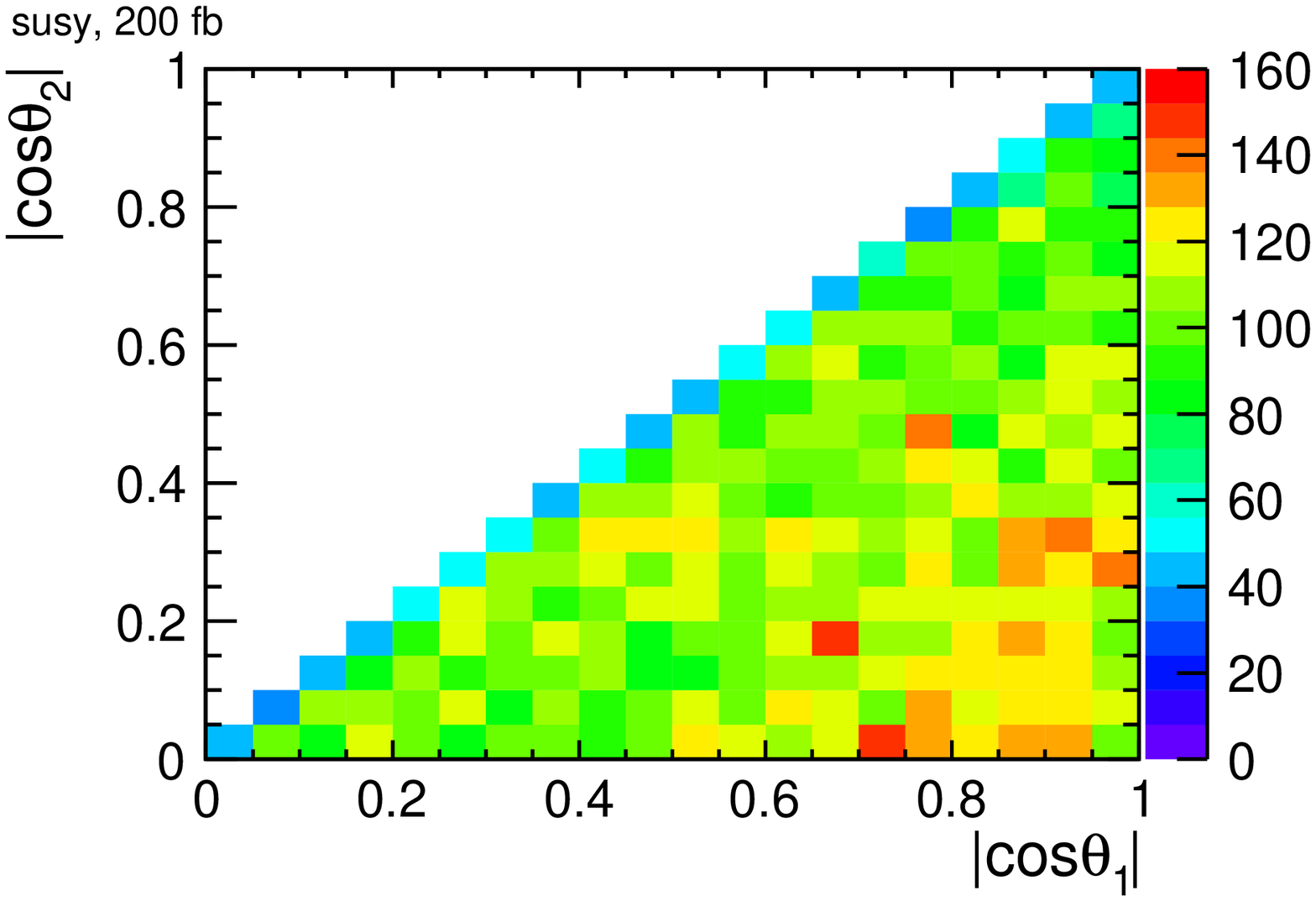}
		(d)
	\end{center}
	\end{minipage}
	\begin{minipage}[t]{.24\textwidth}
	\begin{center}
    \includegraphics[width=1\textwidth]{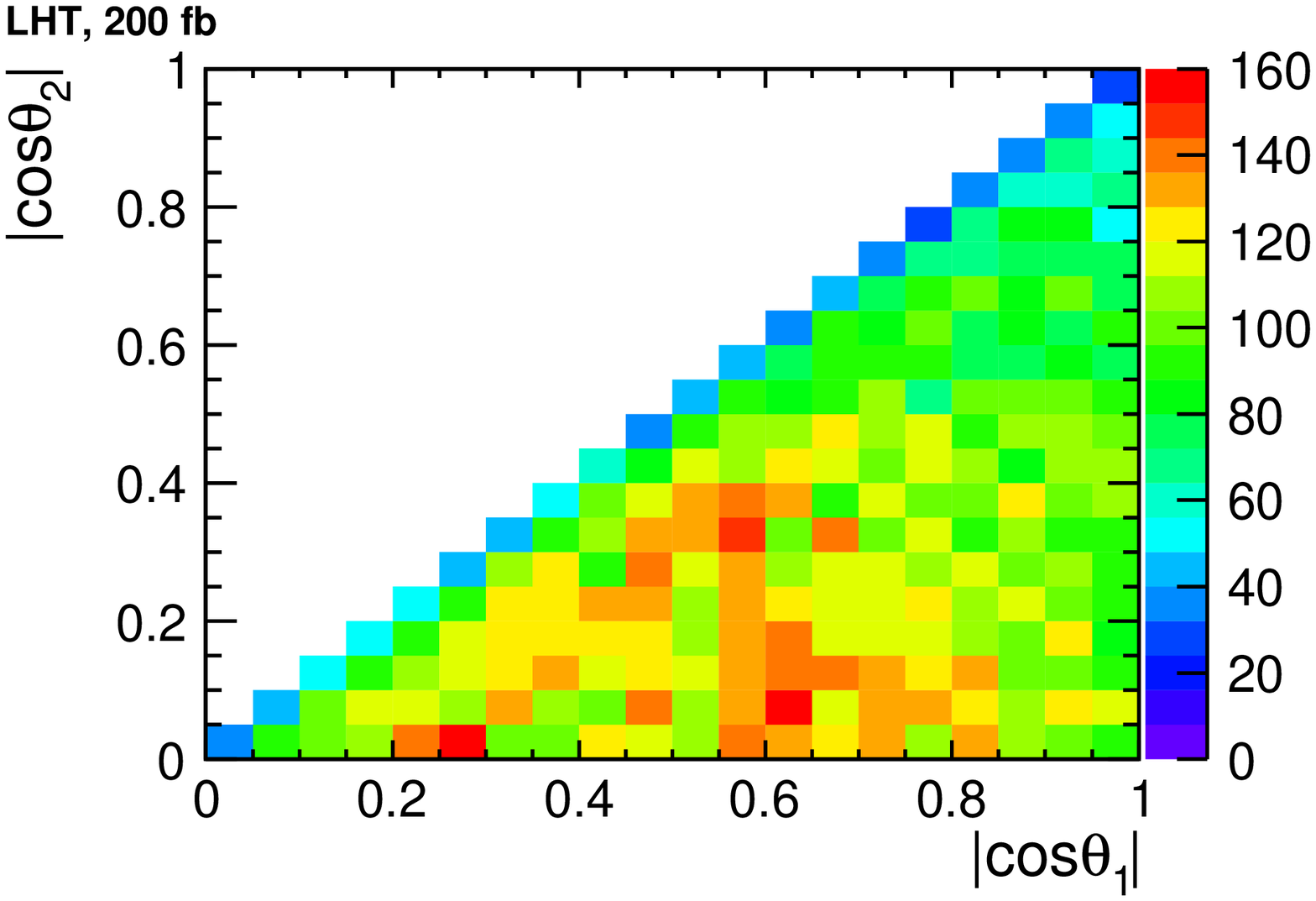}
		(e)
	\end{center}
	\end{minipage}
	\begin{minipage}[t]{.24\textwidth}
	\begin{center}
    \includegraphics[width=1\textwidth]{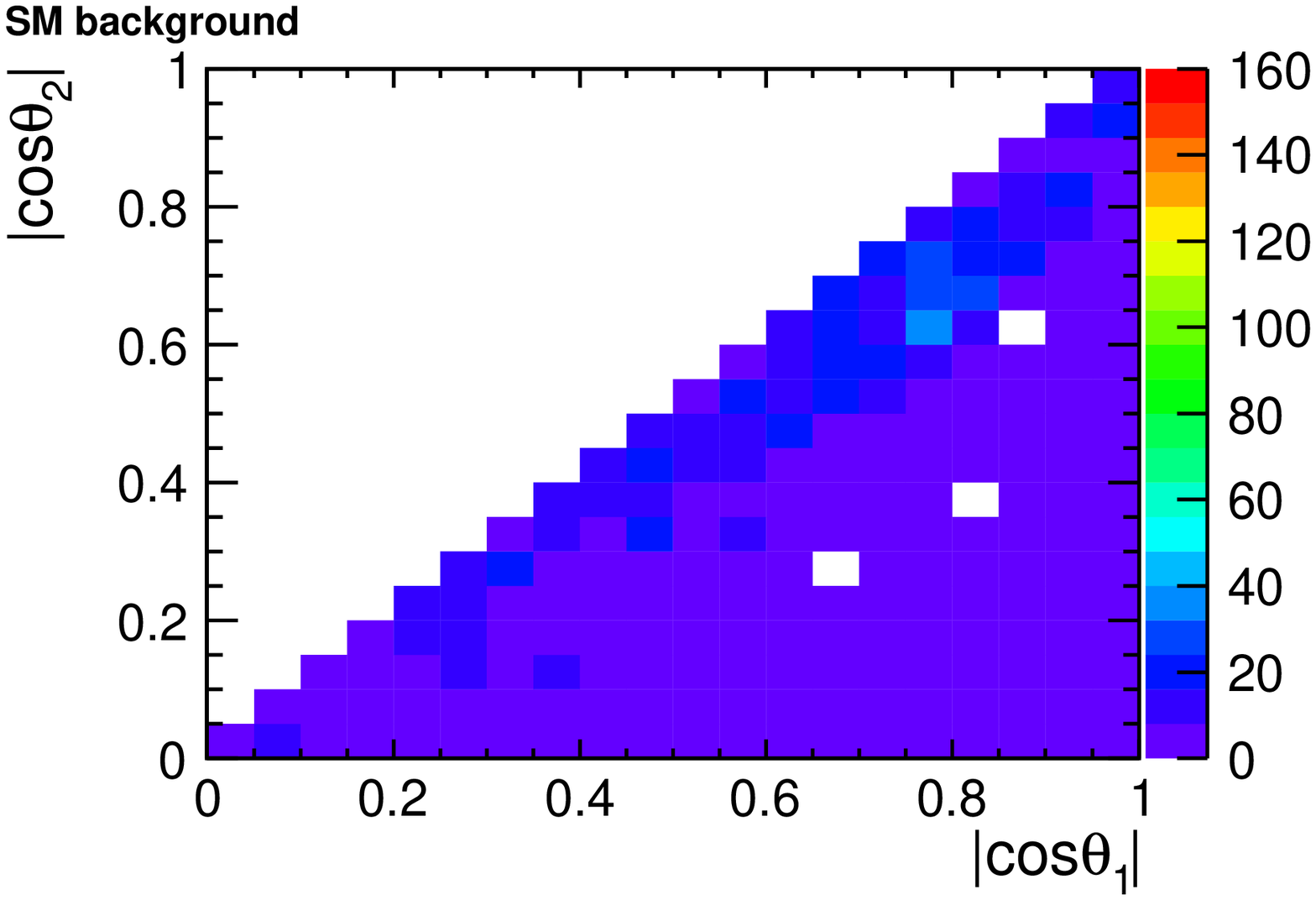}
		(f)
	\end{center}
	\end{minipage}
	\end{center}
	\caption{Production angle distributions with 200 fb cross section.
(a) and (b) show MC and reconstructed 1-dimensional distribution.
Both of two solutions are included.
(c)-(f) give 2-dimensional distribution of IH, SUSY, LH and SM background, respectively.}	
	\label{fig:angle500}
\end{figure}

Separation of the models is possible by comparing distributions of production angles
of the LCPs. To derive the production angles,
a quadratic equation is solved with the masses of new particles
and the momenta of the $W$ bosons with the assumption of back-to-back ejection of two LCPs.
The equation gives either two solutions which contain one correct production angle
or no solutions when the discriminant of the equation is negative.
The unphysical negative discriminant comes from misreconstructing
$W$ momenta or imperfect back-to-back condition of the two LCPs mainly due to initial state radiation.
23.9\% (IH), 20.8\% (SUSY), 23.7\% (LH) and 64.4\% (SM background) of
the events have negative discriminant and are cut off for the following analysis.

Figure \ref{fig:angle500} shows production angle distributions.
1-dimensional results show the visible difference between three models that
IH events concentrate in the central region while SUSY events are almost flatly distributed.
LH events give intermediate profile.
2-dimensional results are actually used to calculate the separation power.
To quantify the difference, we compare the production angle distribution for
one model (dubbed as ``data'') against another model (``template''). 
The chi-square variable $\chi^2$ and the separation power $P$ is given by:
\begin{equation}
	\chi^2 = \sum_i^\mathrm{bins}\frac{(D_i - T_i)^2 - \sigma_{\mathrm{MC},i}^2}{|D_i|}, \qquad P = \frac{\chi^2}{\sqrt{N}}
\end{equation}
where $D_i$ and $T_i$ are number of data and template events (including SM background) in the $i$th bin,
$\sigma_{\mathrm{MC},i}$ is the standard deviation of MC statistics in both data and template samples,
and $N$ is the number of bins (210). The template events are normalized to the integral
of the data events before calculating $\chi^2$.

\begin{table}
	\begin{minipage}[t]{.55\textwidth}
	\begin{center}
	\begin{tabular}{|c|r|r|r|}
	\hline
	Data / Template & IH & SUSY & LH \\\hline\hline
	Inert Higgs			& -  & 86 & 20 \\\hline
	Supersymmetry 	& 71 & -  & 31 \\\hline
	Little Higgs		& 23 & 27 & -  \\\hline
	\end{tabular}
	\\(a) 200 fb
	\end{center}
	\end{minipage}
	\begin{minipage}[t]{.42\textwidth}
	\begin{center}
	\begin{tabular}{|c|r|r|r|}
	\hline
	D / T & IH & SUSY & LH \\\hline\hline
	IH		& -   & 11.7 & 2.6 \\\hline
	SUSY	& 9.9 & -    & 4.5 \\\hline
	LH		& 2.9 &  4.1 & -   \\\hline
	\end{tabular}
	\\(b) 40 fb
	\end{center}
	\end{minipage}
\caption{Separation power $P$ among three models with the 2-dimensional production angle distribution.
The numbers include uncertainty of $\sim2$ for 200 fb and $\sim0.3$
for 40 fb signal cross section due to limited MC statistics.
}
\label{tbl:prodangle}
\end{table}

Table \ref{tbl:prodangle} shows the obtained separation power $P$.
(a) shows enough separation power $> 20 \sigma$ should be obtained with 200 fb signal cross section.
In 40 fb case (b), 2.6 - 2.9 $\sigma$ separation power is expected
between IH and LH models while $> 4 \sigma$ separation is performed in other pairs.

Since the production angle reconstruction uses the masses of the new particles, deviation of mass determination
can cause degradation to the production angle study.
To investigate systematic effect from the mass uncertainty, we produce more template samples with
shifted MC masses of LCP and WIMP. The amount of mass shift is determined as the same as
the mass measurement error at 40 fb signal cross section (shown in Table \ref{tbl:massfit}).
Both negative and positive shifts are considered (four more templates per model are produced).
The separation power obtained by the shifted templates is consistent with the result of the original templates
within the MC statistics.
From that, we can conclude that the effect of the mass uncertainty to the
model separation is no more than the uncertainty of the MC statistics ($\sim2$ in 200 fb and $\sim0.3$ in 40 fb).

\section{Threshold Scan}

\begin{figure}
	\begin{center}
% 	\begin{minipage}[t]{.24\textwidth}
% 	\begin{center}
%     \includegraphics[width=1\textwidth]{rawcs.eps}
% 		(a)
% 	\end{center}
% 	\end{minipage}
	\begin{minipage}[t]{.32\textwidth}
	\begin{center}
    \includegraphics[width=1\textwidth]{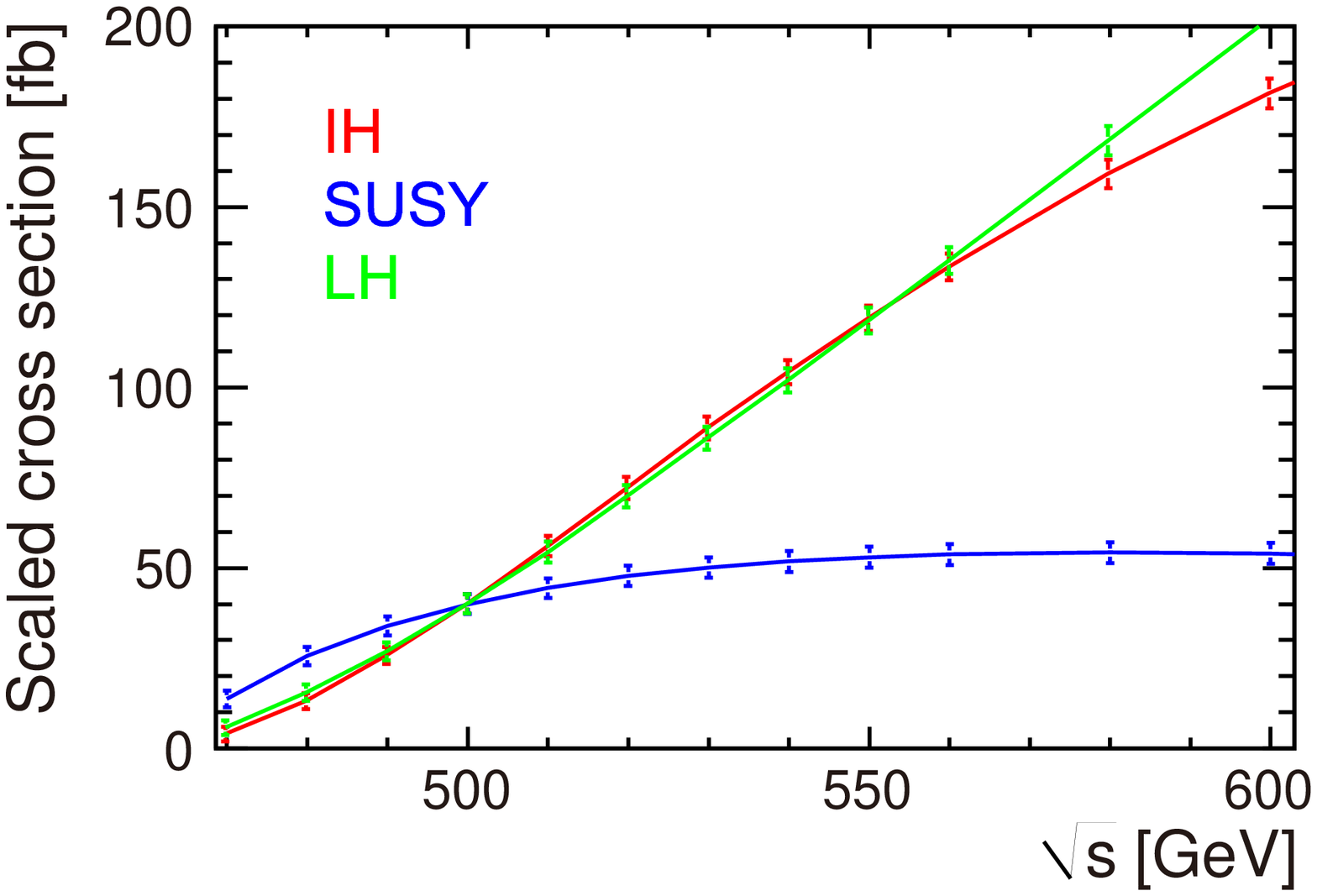}
		(a)
	\end{center}
	\end{minipage}
	\begin{minipage}[t]{.32\textwidth}
	\begin{center}
    \includegraphics[width=1\textwidth]{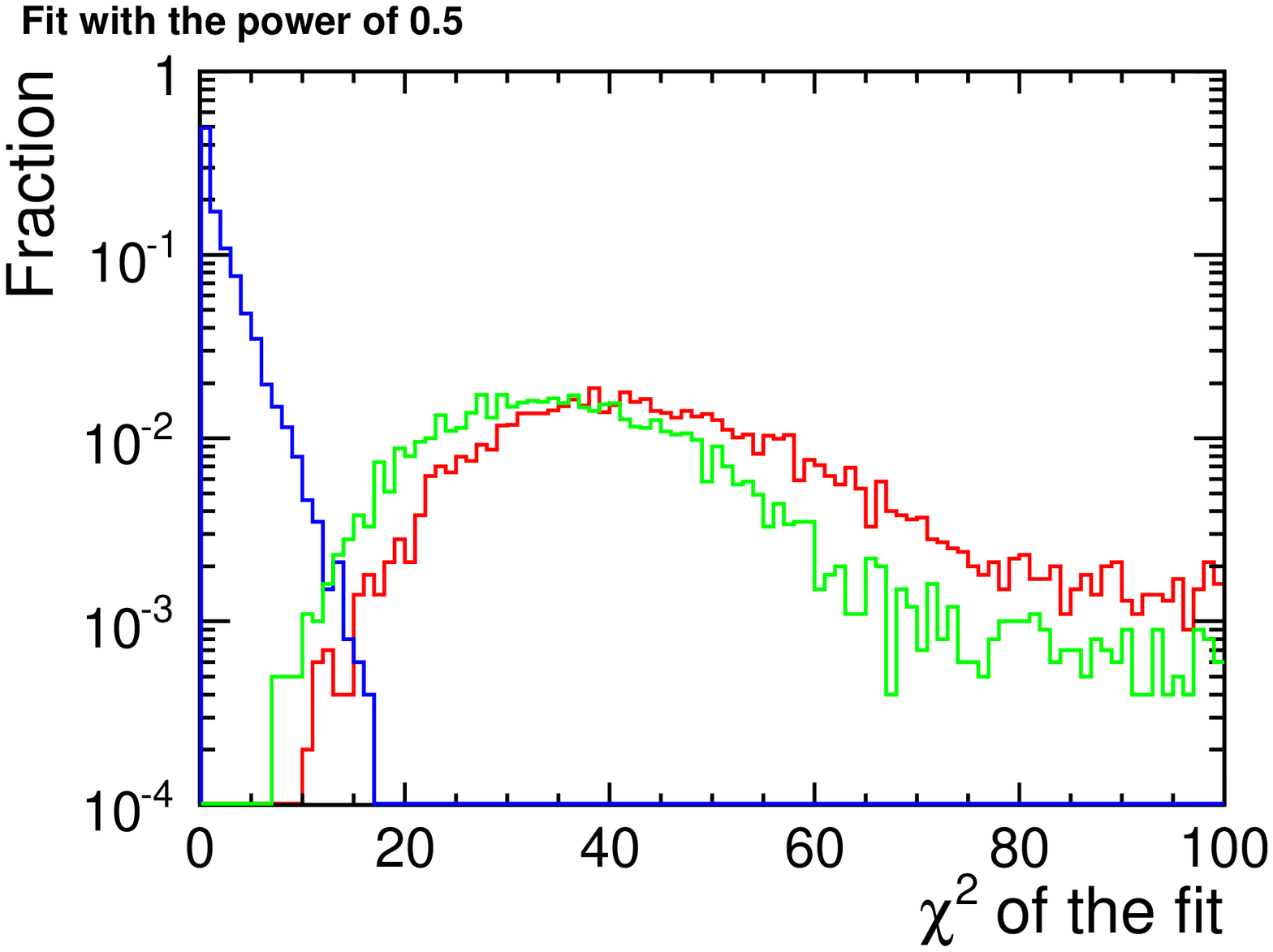}
		(b)
	\end{center}
	\end{minipage}
	\begin{minipage}[t]{.32\textwidth}
	\begin{center}
    \includegraphics[width=1\textwidth]{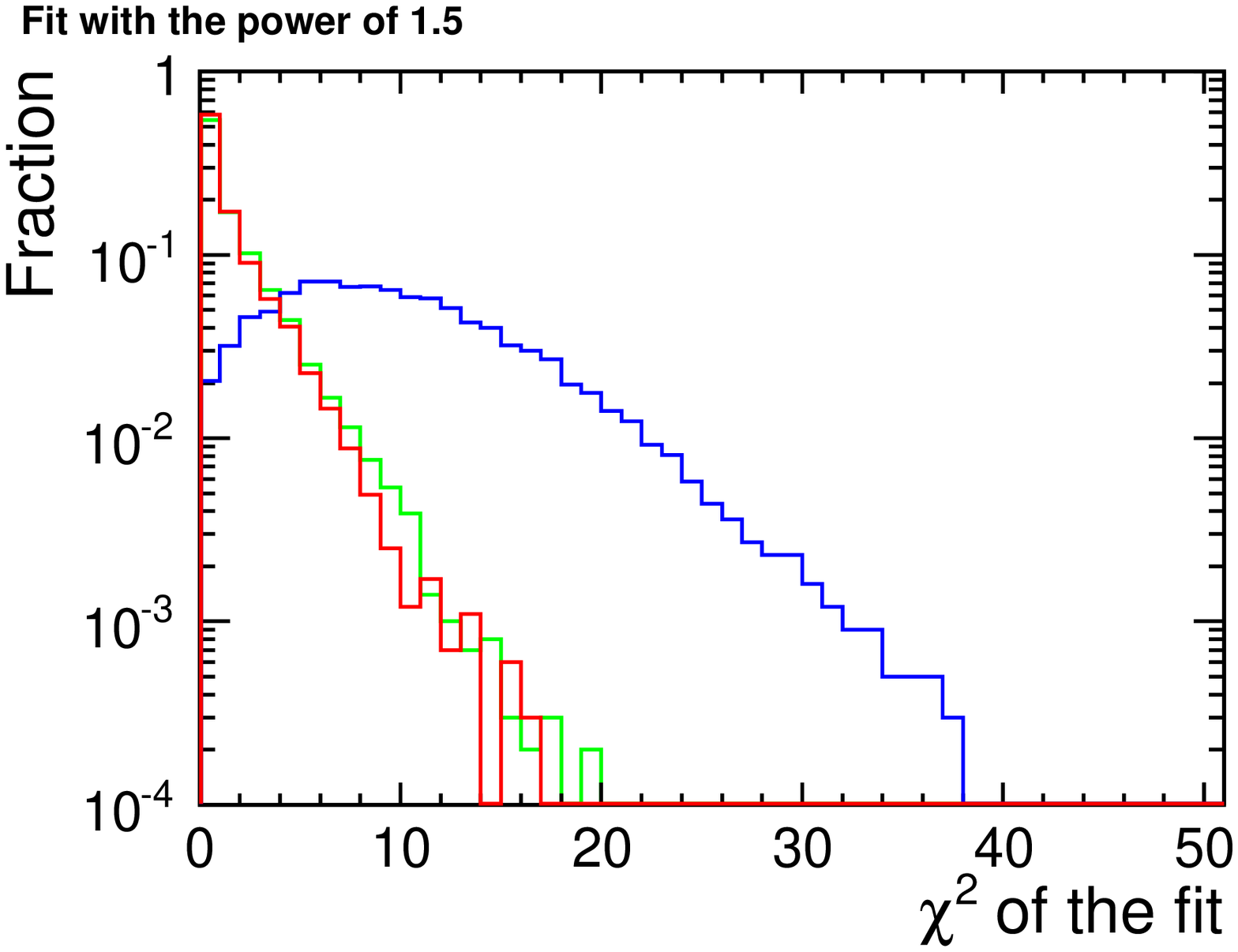}
		(c)
	\end{center}
	\end{minipage}

	\end{center}
	\caption{(a) Dependence of the cross section on the beam energy, normalized to 40 fb at $\sqrt{s} = 500$ GeV.
Error bars are given assuming 50 fb$^{-1}$ data at each point.
 (b)-(c) Result of the $\chi^2$ fit. (b) $(\sqrt{s} - \sqrt{s_0})^{1/2}$ case. (c) $(\sqrt{s} - \sqrt{s_0})^{3/2}$ case.}
	\label{fig:thscan}
\end{figure}

Another strategy to identify the models is to change center-of-mass energy
and monitor difference of the cross section.
Figure \ref{fig:thscan} (a) shows the cross section dependence of each model on $E_\mathrm{CM}$.
A clear difference can be seen as the SUSY model of $E_\mathrm{CM}^{1/2}$ dependence and the other two of $E_\mathrm{CM}^{3/2}$ dependence.

To obtain separation power with the threshold scan, we perform a toy-MC study in which the measured cross section
is fluctuated using signal and background statistics obtained from the full-MC study.
The cut efficiency and background cross section are assumed to be identical to the 500 GeV case in all $\sqrt{s}$ .
We assume a 3-point scan of $\sqrt{s} = $470, 500 and 530 GeV with 50 fb$^{-1}$ integrated luminosity each.
Cross section is set to 40 fb at 500 GeV.

For the separation, we calculate the $\chi^2$ value of the fit of,
\begin{equation}
	\sigma(\sqrt{s}, n) = a(\sqrt{s} - \sqrt{s_0})^{n}, \qquad n = 1/2, 3/2
\end{equation}
where $a$ and $\sqrt{s_0}$ are the free parameters for each model.
Figure \ref{fig:thscan} (b)-(c) shows the $\chi^2$ distributions. With the power of $1/2$ fit (b),
clear separation is obtained between SUSY and other two models. For example, 99.4\% of the SUSY events are within 
$\chi^2 < 12$ while 0.11 and 0.38\% of the IH and LH events remain for the same $\chi^2$ region.
The power of $3/2$ fit (c) has less separation power, and separation between IH and LH is almost impossible by the threshold scan.

\section{Summary}

We study possibility of the identification of WIMP models which have different spin structure
with W + dark matter pair production at $\sqrt{s} = 500$ GeV ILC.
With 200 fb cross section, separation power among three models is strong enough with the production angle distribution
and $\sim 3 \sigma$ separation can still be obtained with 40 fb cross section.
We also show that the threshold scan is a very powerful method for lower statistics, while
in some cases (ex.~IH and LH) behavior around the threshold is similar and separation is not practical.

\section*{Acknowledgments}
The authors would like to thank all members of the ILC physics subgroup for useful discussions.
This study is supported in part by the Creative Scientific Research Grant No.~18GS0202 of the
Japan Society for Promotion of Science.

% ****************************************************************************
% BIBLIOGRAPHY AREA
% ****************************************************************************

\begin{footnotesize}
% IF YOU DO NOT USE BIBTEX, USE THE FOLLOWING SAMPLE SCHEME FOR THE REFERENCES
% ----------------------------------------------------------------------------

% ----------------------------------------------------------------------------

\end{footnotesize}

% ****************************************************************************
% END OF BIBLIOGRAPHY AREA
% ****************************************************************************

\end{document}